# On the Variability of Chaos Indices in Sleep EEG Signals

Amin Banitalebi Dehkordi, Gholam-Ali Hossein-Zadeh

Control and Intelligent Processing Centre of Excellence, School of ECE, College of Engineering, University of Tehran,

banitalebi@ieee.org

***Abstract:*** Previous researches revealed the chaotic and nonlinear nature of EEG signal. In this paper we inspected the variability of chaotic indices of the sleep EEG signal such as largest Lyapunov exponent, mutual information, correlation dimension and minimum embedding dimension among different subjects, conditions and sleep stages. Empirical histograms of these indices are obtained from sleep EEG of 31 subjects, showing that, with a good accuracy, these indices in each stage of sleep vary from healthy human subjects to subjects suspected to have sleep-disordered breathing.

## 1   Introduction

In the previous papers and researches on the human EEG signals, it was shown that human EEG has a chaotic nature. Linear methods of EEG analysis such as Fourier transforms or power spectral analysis, in comparison to chaotic analysis, are more computationally complicated and less strong in the interpreting the results [1, 2]. As another example, [3] shows that using nonlinear time series analysis, we can distinguish even high-dimensional chaos from colored noise. Also some of the traditional linear methods have been found largely insensitive to task conditions associated with different brain dynamics [4, 5].

In this research we show that more than chaotic behaviours, chaotic and nonlinear EEG indices in each stage of sleep vary among: different healthy human subjects; different subjects suspected to have sleep-disordered breathing (Apnea patients); and between these two groups of subjects. Exploring this variability helps the classification and diagnostic studies which is performed via these indices.

In section 2 the important indices are introduced briefly. Section 3 contains a short explanation of the EEG signals which we have used for simulation. In section 4 we state the method of calculation. Results are illustrated in section 5 and section 6 is conclusion.

## 2   Chaotic Indices

In the following subsections we introduce chaotic indices which we used in our research.

### 2.1   Largest Lyapunov Exponent

The Lyapunov Exponent is a number that describes the dynamics of trajectory evolution in the phase space. It capsulizes the average rate of convergence or divergence of two neighbouring trajectories in the phase space. Its value can be negative, zero or positive. Negative values mean that the two trajectories draw closer to one another. Positive exponents on the other hand, result from the trajectory divergence and appear only within the chaotic domain. In other words, a positive Lyapunov exponent is one of the most indicators of the chaos. Here we use the popular Wolf et al. [6] method for our calculation of Largest Lyapunov exponent. This method begins with taking the first point of the reconstructed (lagged) phase space to represent the fiducial trajectory. A nearby point is selected as the first observation along the neighboring trajectory. The gap between two trajectories is monitored over time, until another trajectory becomes closer. At this stage a local slop is computed and the last analyzed point is used for the next reference point. The process is repeated until the fiducial trajectory gets to the end of the data. At this stage it is assumed that the data covers the attractor. Averaging the logs of the absolute values of the various local divergence rates gives the Largest Lyapunov Exponent, as the following equation:

$$l_1 = \lim_{n \to \infty} \frac{1}{n} \sum_{i=0}^{n-1} \log_e |f'(x_i)| \qquad (1)$$

In this equation, $f(x_i)$ is the local slop mentioned above. Because it averages local divergences and/or convergences from many places over the entire attractor, a *Lyapunov Exponent is a global quantity*, not a local quantity. The largest Lyapunov Exponent can be interpreted in three ways:

- In *n* dimensions, $l_1$ quantifies, in a single number, the average rate at which the fastest growing phase space dimension grows.
- It quantifies the average predictability over the attractor.
- Because the neighboring trajectories represent changes in initial conditions of a system, $l_1$ is an average or global measure of the sensitivity of the system to slight changes or perturbations. A system isn't sensitive at all in the nonchaotic regime, since any two nearby trajectories converge. In contrast, a system is highly sensitive in the chaotic regime, in that two neighboring trajectories separates, sometimes rapidly.

### 2.2 Mutual Information

If we denote $X, Y$ as two random variables, then $H_X, H_Y$ are their entropies and we have:

$$H_Y = -\sum_{j=1}^{N_S} p(y_j) \log_2 p(y_j) \qquad (2)$$

in which $N_s$ is the number of non-zero probabilities.

The Mutual Information for $X, Y$ is defined as:

$$I_{Y;X} = H_Y + H_X - H_{X;Y} \qquad (3)$$

in which $H_{X;Y}$ is defined as:

$$H_{X;Y} = -\sum_{i=1}^{N_S} \sum_{j=1}^{N_S} p(x_i, y_j) \log_2 p(x_i, y_j) \qquad (4)$$

After substituting the above Eq. in (3) and by some mathematical simplifications, we will have:

$$I_{Y;X} = \sum_{i=1}^{N_S} \sum_{j=1}^{N_S} p(x_i, y_j) \log_2 \frac{p(x_i, y_j)}{p(x_i) p(y_j)} \qquad (5)$$

For the calculation of $I_{Y;X}$ in EEG signals, in lag space $x_i$ becomes $x_t$ and $x_j$ becomes $x_{t+m}$. Bigger quantity of mutual information results in a less chaotic system. More details are reported in [7].

### 2.3 Minimum Embedding Dimension

For computational costs, simplicity of interpretation and other reasons, we'd like to reconstruct an attractor in a *small* embedding dimension. (After the attractor is reconstructed, larger dimensions also suffice, but we want the minimum possible.) There's no theory or even a rule-of-thumb available in this regard. None of the many proposed ways to estimate the minimum embedding dimension is yet widely accepted. To solve the problem of false neighbor method, Cao proposed a method to choose the threshold value, which is often used to determine the embedding dimension. Different time series data may have different threshold values. These imply that it is difficult to give an independent reasonable threshold value which is independent of the dimension *d* and each trajectories point, as well as the considered time series data. In this method a new quantity is defined:

$$E(m) = \frac{1}{N-mt} \sum_{i=1}^{N-mt} f(i,m) \qquad (6)$$

$E(m)$ is dependent only on the dimension *m* and the lag *t* and *f* is a function of $i, m$. To investigate its variation from *m to m+1*, *E1(m)* is defined as:

$$E1(m) = \frac{E(m+1)}{E(m)} \qquad (7)$$

Cao found that *E1(m)* stops changing when *m* is greater than some value $m_0$ if the time series comes from an attractor. Then

$m_0 +1$ is the minimum embedding dimension we look for. It is necessary to define another quantity which is useful to distinguish between deterministic and stochastic signals. Let

$$E2(m) = \frac{E^*(m+1)}{E^*(m)} \quad (8)$$

where:

$$E^*(m) = \frac{1}{N-mt} \sum_{i=1}^{N-mt} |x_{i+mt} - x_{n(i,m)+mt}| \quad (9)$$

For time series data from a random set of numbers, $E1(m)$, in principle, will never attain a saturation. More details are in [8].

**2.4 Correlation Dimension**

The correlation dimension is the most popular non integer dimension currently used. It probes the attractor to a much finer scale than does the information dimension (see [7] for more details about information dimension) and is also easier and faster to compute. Like the information dimension, it takes into account the frequency with which the system visits different phase space zones. Most other dimensions involve moving a measuring device by equal, incremental lengths over the attractor (tantamount to placing a uniform grid over it). In contrast, the correlation dimension involves systematically locating the device at each datum point, in turn. The procedure usually begins by embedding the data in a two-dimensional pseudo phase space. For a given radius $e$, count the number of points within distance $e$ from the reference point. After doing that for each point on the trajectory, sum the counts and normalize the sum. That yields a correlation sum. Then repeat that procedure to get correlation sums for larger and larger values of $e$. A log plot of correlation sum versus $e$ (for that particular embedding dimension) typically shows a straight or nearly straight central region. The slop of that straight segment is the correlation dimension. The next step is to repeat the entire procedure for larger and larger embedding dimensions. For chaotic data, the correlation dimension initially increases with embedding dimension, but eventually (at least in the ideal case) it asymptotically approaches a final (true) value. Grassberger and Procaccia's [9] method for computing the correlation dimension $D_2$ is mathematically illustrated below: First of all, a phase space must be constructed. This space should being spanned by a set of embedding vectors, in the case of univariate signals, following a proposal made by Takens [10], called the time shift method, n-dimensional vectors are constructed in the following way:

$$\vec{x} = \{x(t), x(t+t), ..., x(t+(n-1)t)\} \quad (10)$$

$t$, is a fixed time increment in above Eq. and n is the embedding dimension. Every instantaneous state of the system is therefore is represented by the vector $\vec{x}$ which defines a point in phase space. Once the phase space is constructed, the correlation integral as a function of variable distances $R$ is defined as:

$$C(R) = \lim_{n \to \infty} \frac{1}{N^2} \sum_{i \neq j} \Theta(R - |\vec{x}_i - \vec{x}_j|) \quad (11)$$

Where $N$ is the number of data points and $\Theta$ is the Heaviside function (Heaviside function is zero if its argument is negative; and 1 if the argument is zero or positive). Thus $C(R)$ is the probability that two arbitrary points $\vec{x}_i, \vec{x}_j$ will be separated by a distance less than $R$. Theiler (1986) made a correction to this method in order to avoid spurious temporal correlations. He proposed that the vectors to be compared when calculating the correlation integral, should be distanced at least $W$ data points ($|i-j|>W$), where $W$ is a measure of temporal correlation of the signal (e.g. the first zero of the autocorrelation function). In the case the attractor is a simple curve in the phase space, the number of pair of vectors whose distance is less than a certain radius $R$ will be proportional to $R^1$. In the case the attractor is a two dimensional surface, $C(R) \sim R^2$; and for a fixed point, $C(R) \sim R^0$. Generalizing we can write the following relation:

$$C(R) \sim R^{D_2} \quad (12)$$

thus, if the number of data and the embedding dimension are sufficiently large we obtain

$$D_2 = \lim_{R \to 0} \frac{\log(C(R))}{\log(R)} \quad (13)$$

By plotting $\log(C(R))$ versus $\log(R)$, $D_2$ can be calculated from the slop of the curve. More details are in [11].

**3 The EEG Data**

EEG signals were obtained from Physionet databank, both for healthy and abnormal subjects [12]. There were 6 healthy subjects and 25 Apnea patients. Each of Healthy EEG's is a 24 hour recorded signal with a sampling frequency of 100 Hz, while each of the abnormal EEG signals is an about of 6 hour signal, recorded at sampling frequency of 128 Hz.

It is noteworthy that a pre-processing stage was done on the dataset to remove artifacts mainly caused by EOG, to reach a better signal for other processing levels. To do this we use the *HinfTV regression* algorithm which is described in [18].

## 4  Method

In the analysis of each signal, each window of the signal was 30 seconds wide. This length is equal to one (minimum) separate stage of sleep. The standard rule of annotating was used for sleep stages (Wake, REM, Stage1, Stage2, Stage3, and Stage4). For each stage of the sleep, both in normal subjects and abnormal ones, the chaos indices including largest Lyapunov exponent, mutual information, correlation dimension and minimum embedding dimension was calculated for all (windows of) signals. To simplify the calculations twelve signals were synthesized by concatenation of similar stages of healthy (patient) subjects together (2 groups in 6 stages). Each signal corresponds to one stage of sleep, either in healthy or patient subjects.

## 5  Results

Table 1, summarizes the mean and standard deviation (STD) of various indices in different stages of sleep, among healthy and abnormal subjects. Although some of indices seem to be different between healthy and abnormal subjects, but one must consider the variability of them inside each group.

Table 1: Variability of Chaos Indices in Four Sleep Stages, between Healthy and Abnormal Subjects

|  | Stage 1 | | Stage 2 | | Stage 3 | | Stage 4 | |
| --- | --- | --- | --- | --- | --- | --- | --- | --- |
|  | Healthy | Apnea | Healthy | Apnea | Healthy | Apnea | Healthy | Apnea |
|  | mean (STD) | mean (STD) | mean (STD) | mean (STD) | mean (STD) | mean (STD) | mean (STD) | mean (STD) |
| largest Lyapunov exponent | 8.8919 (0.1046) | 9.1855 (0.2238) | 8.8868 (0.0972) | 9.1771 (0.1889) | 8.8894 (0.0871) | 9.2011 (0.1432) | 8.8566 (0.1317) | 9.1986 (0.1014) |
| mutual information | 3.0152 (0.0235) | 2.7417 (0.0854) | 3.0189 (0.0212) | 2.7531 (0.0746) | 3.0192 (0.0178) | 2.7627 (0.0647) | 3.0189 (0.0306) | 2.7758 (0.0482) |
| minimum embedding dimension | 0.6276 (0.0118) | 0.6229 (0.0235) | 0.6279 (0.0035) | 0.6246 (0.0069) | 0.6283 (0.0033) | 0.6284 (0.0065) | 0.6288 (0.0041) | 0.6328 (0.0062) |
| correlation dimension | 2.8788 (1.2839) | 3.7006 (1.4040) | 3.1287 (1.3017) | 3.2993 (1.2381) | 3.3172 (1.3222) | 3.7693 (1.1469) | 2.5722 (1.4901) | 4.1949 (1.0908) |

As a complete illustration, Fig. 1 shows the relative frequency of the largest Lyapunov exponent in normal and abnormal states in stage 1 (Both are normalized). The curve with dots corresponds to healthy subjects. Thus these curves may be interpreted as probability density functions (pdf) of largest Lyapunov exponent for healthy and abnormal subjects. Fig. 2 compares mutual information in stage 1 between the two states, Fig. 3 does the same for correlation dimension and Fig. 4 compares minimum embedding dimension between them, another time in stage 1. In some other stages, events were more separable and in some were less distinguishable. Almost in all indices in the wake stage, indices were not sensitively different, but in stage 4 we had the best results.

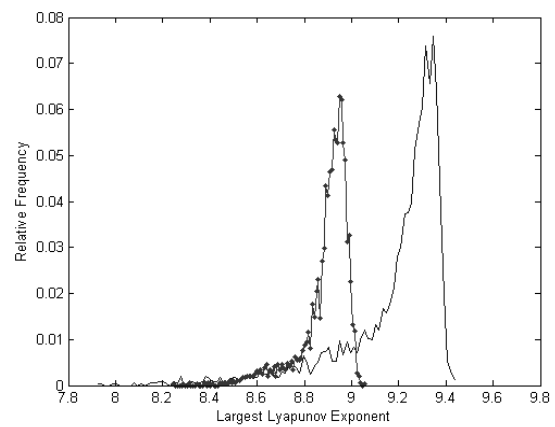

Fig. 1: Comparison of Largest Lyapunov Exponent in Stage 1

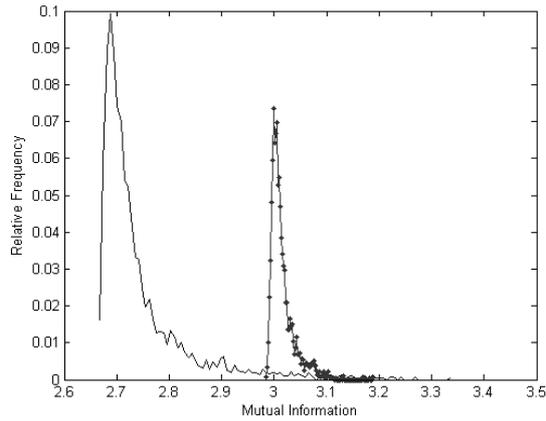

Fig. 2: Comparison of Mutual Information in Stage 1

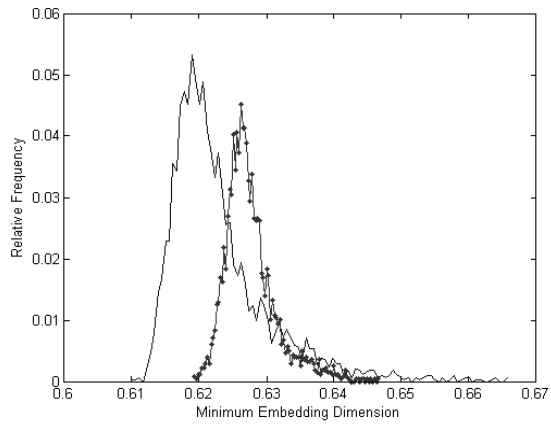

Fig. 3: Comparison of Minimum Embedding Dimension in Stage 1

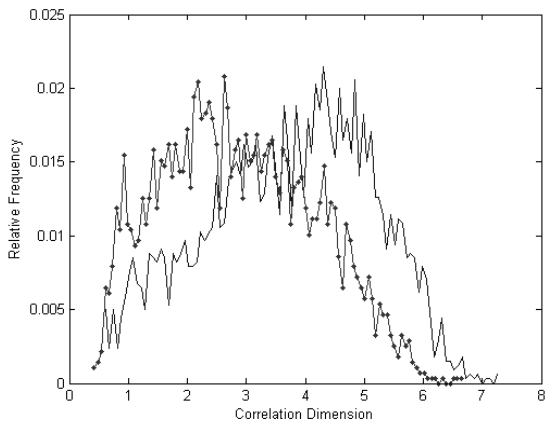

Fig. 4: Comparison of Correlation Dimension in Stage 1

In order to verify that in the presence of intragroup variability of each index, whether it still distinguish healthy and abnormal subjects, we used a standard t-test measure. Without loss of generality if an index has mean $m_1$ and standard deviation of $s_1$ in group 1, and mean $m_2$ and standard deviation of $s_2$ in group 2, then a t-value can be assigned to it as:

$$T = \frac{m_1 - m_2}{\sqrt{\frac{s_1^2}{n_1} + \frac{s_2^2}{n_2}}} \quad (14)$$

in which $n_1, n_2$ are the sample sizes of group 1 and 2 respectively.

A p-value can be assigned to each T that indicates how much error we would have in separation between group1 and group2 using the mentioned index. The p value is calculated as: $p = 1 - CDF(T)$, in which CDF(.) is the cumulative distribution function of t-student. Thus a small p-value shows a good reparability and vice-versa. Further information about p-values can be achieved in [13]. Tables 2-5 show the p-value of each chaos index in separating each stage between healthy and patient subjects.

Table 2: P-values For Largest Lyapunov Exponent, as a Measure of Error Percentage in Separation between Healthy and Abnormal Subjects

| Stage | p-value |
|---|---|
| Wake | 0.02 |
| REM | 0.0005 |
| Stage 1 | 0.0005 |
| Stage 2 | 0.0005 |
| Stage 3 | 0.0005 |
| Stage 4 | 0.0005 |

Table 3: P-values for Mutual Information, as a Measure of Error Percentage in Separation between Healthy and Abnormal Subjects

| Stage | p-value |
|---|---|
| Wake | 0.0005 |
| REM | 0.0005 |
| Stage 1 | 0.0005 |
| Stage 2 | 0.0005 |
| Stage 3 | 0.0005 |
| Stage 4 | 0.0005 |

Table 4: P-values for Minimum Embedding Dimension, as a Measure of Error Percentage in Separation between Healthy and Abnormal Subjects

| Stage   | p-value |
|---------|---------|
| Wake    | 0.0005  |
| REM     | 0.0005  |
| Stage 1 | 0.0005  |
| Stage 2 | 0.0005  |
| Stage 3 | 0       |
| Stage 4 | 0.0005  |

Table 5: P-values for Correlation Dimension, as a Measure of Error Percentage in Separation between Healthy and Abnormal Subjects

| Stage   | p-value |
|---------|---------|
| Wake    | 0.0005  |
| REM     | 0.0005  |
| Stage 1 | 0.0025  |
| Stage 2 | 0.0005  |
| Stage 3 | 0.0005  |
| Stage 4 | 0.0005  |

As we see from the above tables and diagrams, chaos indices mentioned above, can approximately separate different stages of sleep in healthy and abnormal subjects.

## 6   Conclusion

In this paper we calculated several indices of chaos for sleep EEG signals of 6 healthy and 25 patients (Apnea). We explored the variability of indices among different subjects, and evaluated the separability of them in two groups. In conclusion these indices can approximately separate different stages of sleep in healthy and abnormal subjects.